# Well-balanced and flexible morphological modeling of swash hydrodynamics and sediment transport

by Peng HU[a, b], Wei LI[a], Zhiguo HE [a, b, *], Thomas Pähtz[a, b], Zhiyuan YUE[c]

[a] Ocean College, Zhejiang University, Hangzhou, China;

[b] State Key Laboratory of Satellite Ocean Environment Dynamics, The Second Institute of Oceanography, Hangzhou, China;

[c] Changjiang Waterway planning Design and Research Institute, Wuhan, China.

* Corresponding author

Emails: pengphu@zju.edu.cn (Peng Hu), lw05@zju.edu.cn (Wei Li), hezhiguo@zju.edu.cn (Zhiguo He)

**Abstract**: Existing numerical models of the swash zone are relatively inflexible in dealing with sediment transport due to a high dependence of the deployed numerical schemes on empirical sediment transport relations. Moreover, these models are usually not well-balanced, meaning they are unable to correctly simulate quiescent flow. Here a well-balanced and flexible morphological model for the swash zone is presented. The nonlinear shallow water equations and the Exner equation are discretized by the shock-capturing finite volume method, in which the numerical flux and the bed slope source term are estimated by a well-balanced version of the SLIC (Slope LImited Centered) scheme that does not depend on empirical sediment transport relations. The satisfaction of the well-balanced property is demonstrated through simulating quiescent coastal flow. The quantitative accuracy of the model in reproducing key parameters (i.e., the notional shoreline position, the swash depth, the flow velocity, the overtopping flow volume, the beach change depth and the sediment transport rate) is shown to be satisfactory through comparisons against analytical solutions, experimental data as well as previous numerical solutions. This work facilitates an improved modeling framework for the swash hydrodynamics and sediment transport.

## 1. Introduction

The term "swash zone" refers to narrow coastal regions that are successively covered and uncovered by water due to run-up and backwash of waves (Masellink and Hughes 1998; Masellink and Puleo 2006). Although the swash hydrodynamics and sediment transport have been studied for three decades (Brocchini 2006), they still are insufficiently understood. For example, in a very recent paper by van Rijn (2014), the challenge of estimating the sediment transport rate in the swash zone was mentioned at least seven times. Here we present a morphological modeling technique for the swash zone which is both "well-balanced" and "flexible", the meaning of which is explained in the following paragraphs.

Among the different modeling approaches (e.g., the Reynolds-averaged Navier-Stokes equations, the Boussinesq equations, and the nonlinear shallow water equations), the present model adopts the nonlinear shallow water equations due to its relative simplicity but widely recognized applicability in the swash zone (Hibberd and Peregrine 1979; Brocchini and Dodd 2008). As suggested in Toro (2001) and Brocchini and Dodd (2008), the total variation diminishing (TVD) finite volume method, which is shock-capturing, is one of the most promising methods for solving the nonlinear shallow water equations. According to this method, the finite volume discretization is implemented to solve the governing equations, and the numerical flux is treated as a Riemann problem. Examples for modeling coastal flows are abundant, such as those using the exact Riemann solver (Wei *et al.* 2006), the weighted average flux (WAF) solver (Brocchini *et al* 2001; O'Donoghue *et al.* 2010; Postacchini *et al.* 2012, 2014), the approximate Roe's solver (Dodd 1998; Hubbard and Dodd 2002), the HLL solver (Hu *et al.* 2000; Borthwick *et al.* 2006; Briganti and Dodd 2009a, b; Mahdavi and Talebbeydokhti 2011; Kuiry *et al.* 2012), and the centered approximate Riemann solver (Mahdavi and Talebbeydokhti 2009). Other models for the swash zone include those based on the method of characteristics (Kelly and Dodd 2010; Zhu *et al.* 2012; Zhu and Dodd 2013) and the McCormack method (Briganti *et al.* 2012a, b), etc. Great progress has been made, but further improvement is critical for refined modeling quality, as detailed below.

First, as discussed in the comprehensive review by Brocchini and Dodd (2008), most numerical models have focused on the swash flow free of sediment transport (Dodd 1998; Hu *et al.* 2000; Hubbard and Dodd 2002; Borthwick *et al.* 2006; Wei *et al.* 2006; Brocchini *et al.* 2001; Briganti and Dodd 2009a, b; Mahdavi and Talebbeydokhti 2009, 2011; O'Donoghue *et al.* 2010; Kuiry *et al.* 2012), whereas only a few morphological models have been developed for the swash zone (Masselink and Li 2001; Dodd *et al*. 2008; Kelly and Dodd 2010; Briganti *et al.* 2012a, b; Postacchini *et al*. 2012, 2014; Zhu *et al.* 2012; Zhu and Dodd 2013). However, due to the dominant use of the upwind-type numerical schemes and thus the involvement of the eigenstructures of the governing equations that depend on the use of a particular empirical sediment transport relation, it can be quite tedious to modify these models to implement other sediment transport relations, which makes them relatively inflexible. Unfortunately, flexibility in dealing with sediment transport is quite important due to the fact that there is no generally applicable empirical sediment transport relation, nor is it likely that any will be available in the near future. It is therefore no surprise to see the development of a TVD-McCormack scheme-based morphological model by Briganti *et al.* (2012a, b) that was designed to be flexible. However, an empirical sediment relation is still involved in the TVD-McCormack scheme. Moreover, the necessity of switching off the TVD function for sub-threshold sediment motion makes the model difficult to be extended to non-uniform sediment transport because sediment particles of different sizes have different threshold conditions (Hu et al. 2014). Note that the centered-type finite volume method intrinsically avoids the use of the aforementioned eigenstructures (Toro et al. 2009; Canestrelli et al. 2010), and thus allows for a more flexible modeling regarding the use of empirical sediment transport equations. However, this modeling technique has rarely been seen in morphological modeling of coastal flows (Mahdavi and Talebbeydokhti 2009).

Second, existing numerical models for the swash zone rarely consider the well-balanced property ( i.e., the balance between the bed slope source term and the flux gradient), making them in most cases unable to simulate quiescent flow (Hubbard and Dodd 2002; Wei et al. 2006; Mahdavi and Talebbeydokhti 2009). The reason for why so little of attention has been paid to the well-balanced property is that most coastal flows are highly dynamic (Brocchini and Dodd 2008). However, the inability to model quiescent flows necessarily introduces uncertainties even to the modeling of such highly dynamic flows. It is the unknown magnitude of these uncertainties, which makes well-balanced modeling highly desirable (Zhou et al. 2001; Hubbard and Dodd 2002; Wei et al. 2006; Aureli et al. 2008; Hu et al. 2012; Mahdavi and Talebbeydokhti 2009; Canestrelli et al. 2010; Donat et al. 2014; Bollermann et al. 2013; Capilla and Balaguer-Beser 2013; Hou et al. 2013; Siviglia et al. 2013; Li et al. 2014).

Motivated by the above background, this paper presents a well-balanced and flexible morphological model for coastal engineering purposes and tests its quantitative accuracy for swash flow. The governing equations of the model include the nonlinear shallow water equations for swash hydrodynamics and the Exner equation for beach morphological change. The governing equations are discretized by the shock-capturing finite volume method, in which the numerical flux and the bed slope source term are estimated by a well-balanced version of the SLIC (slope limited centered) scheme. The model is thoroughly validated through comparisons against analytical solutions, existing numerical solutions, and experimental data. Particular attention is paid to quantitative accuracy of the shoreline position, the swash depth, the swash velocity, the overtopping flow volume, the beach morphological change, and the sediment transport rate. The contribution of this paper is two-fold. First, it may be one of the first well-balanced morphological models for the swash zone. Second, it facilitates a flexible morphological modeling framework for the swash zone.

## 2. Mathematical formulations

### 2.1 Governing equations

The nonlinear shallow water equations coupled with the Exner equation for the swash zone are written in a vector form as (Kelly and Dodd 2010; Briganti *et al.* 2012a, b; Postacchini *et al.* 2012, 2014; Zhu *et al.* 2012; Zhu and Dodd 2013)

$$\frac{\partial \mathbf{U}}{\partial t} + \frac{\partial \mathbf{F}}{\partial x} = \mathbf{S}_b + \mathbf{S}_f \qquad (1)$$

$$\mathbf{U} = \begin{bmatrix} h \\ hu \\ (1-p)z \end{bmatrix}, \quad \mathbf{F} = \begin{bmatrix} hu \\ hu^2 + gh^2/2 \\ q_b \end{bmatrix}, \quad \mathbf{S}_b = \begin{bmatrix} 0 \\ (-\partial z/\partial x)gh \\ 0 \end{bmatrix}, \quad \mathbf{S}_f = \begin{bmatrix} 0 \\ -\tau_b/\rho_w \\ 0 \end{bmatrix} \qquad (2a, b, c, d)$$

where $t$ = time; $x$ = horizontal coordinate; $h$ = flow depth, $u$ = depth-averaged flow velocity, $p$ = bed sediment porosity, $z$ = beach elevation, $g$ = gravitational acceleration; $q_b$ = sediment transport rate, $-\partial z/\partial x$ = beach slope, $\tau_b$ = bed shear stress, and $\rho_w$ = density of water. Here the bed shear stress is estimated as a bulk frictional force using the Manning roughness: $\tau_b = \rho_w u_*^2 = \rho_w g n^2 u|u|/h^{1/3}$, where $n$ = Manning roughness, $u_*$ = bed shear velocity. The empirical relation for the sediment transport rate will be introduced in relation to the specific numerical cases in Section 3.

### 2.2 Numerical scheme

*2.2.1 Finite volume discretization*

Implementing the finite volume discretization along with the operator-splitting method, one obtains (Aureli *et al.* 2008; Hu *et al.* 2012)

$$\mathbf{U}_i^* = \mathbf{U}_i^n - \frac{\Delta t}{\Delta x}[\mathbf{F}_{i+1/2}^n - \mathbf{F}_{i-1/2}^n] + \Delta t \mathbf{S}_{bi} \qquad (3a)$$

$$\mathbf{U}_i^{n+1} = \mathbf{U}_i^* + \Delta t \mathbf{S}_f(\mathbf{U}_i^*) \qquad (3b)$$

where $\Delta x$ = spatial step, $\Delta t$ = time step, the superscripts $n$ and $*$ are time step indexes, the subscript $i$ = spatial index, and $\mathbf{F}_{i+1/2}$ = inter-cell numerical flux. The time step is constrained by the Courant-Friedrichs-Lewy condition using the Courant number $Cr$ as a controller: $Cr = (u + \sqrt{gh})\Delta t/\Delta x \leq 1$.

*2.2.2 Estimation of the numerical flux and bed slope source term*

This sub-section introduces the WSDGM (weighted surface depth gradient method) version of the SLIC scheme, which is well-balanced and used to estimate the numerical flux and bed slope source term. The SLIC scheme results from the replacement of the Godunov flux (an upwind Riemann solver) by the first order centered (FORCE) flux in the MUSCL-Hancock scheme (Toro 2001). In the SLIC scheme, the estimation of the numerical flux is seen as a Riemann problem with the two edge states defined at the two sides (referred to as left side and right side below) of the inter-cell edge. To achieve the second-order accuracy, the edge states are obtained by first using the interpolation from the cell center to the cell edge, and second evolving the interpolated states over a half time step. The original SLIC scheme is termed as the DGM (depth gradient method) version because it makes use of the spatial gradient of the water depth for the interpolation, which is stable for cases with small spatial gradients of the water depth. In practice, bed topographies are usually irregular and favor large spatial gradients, for which the DGM version may not be able to preserve the quiescent flow because of the unbalance between the bed slope source term and the flux gradient over an irregular topography. This motivated the development of the SGM (surface gradient method) by Zhou *et al.* (2001), which is well-balanced but may produce physically unrealistic results over a regular topography (Aureli *et al.* 2008). Retaining the good capabilities of both the DGM and the SGM, Aureli *et al.* (2008) developed the WSDGM, which has been extended to modeling subaqueous turbidity currents (Hu *et al.* 2012). The WSDGM version of the SLIC is adapted for the swash zone.

***Step 1:*** *Spatial interpolation for the second-order accuracy*

A new vector $\mathbf{Q} = [0,\ hu,\ (1-p)z,\ h,\ (h+z)]^T$ is introduced for a convenient mathematical description, where the superscript $T$ represents the mathematical operation of matrix transpose. The first element of the new vector $\mathbf{Q}$ is purposely set to zero, so that a weighted value between the DGM and the SGM can be prescribed to it, see below Eq. (7a, b). The edge variables $\mathbf{Q}^L_{i+1/2}$ and $\mathbf{Q}^R_{i+1/2}$ at the left and right sides of $x_{i+1/2}$ are interpolated from the cell-center variables, which read

$$\mathbf{Q}^L_{i+1/2} = \mathbf{Q}^n_i + \varphi^L_{i-1/2} \frac{\mathbf{Q}^n_{i,j} - \mathbf{Q}^n_{i-1}}{2} \tag{4a}$$

$$\mathbf{Q}^R_{i+1/2} = \mathbf{Q}^n_{i+1} - \varphi^R_{i+3/2} \frac{\mathbf{Q}^n_{i+2} - \mathbf{Q}^n_{i+1}}{2} \tag{4b}$$

where $\varphi$ = slope limiter. Here the vanLeer limiter is used, which reads

$$\varphi(\mathbf{r}) = (\mathbf{r} + |\mathbf{r}|)/(1+\mathbf{r}) \tag{5}$$

with

$$\mathbf{r}^L_{i-1/2} = \frac{\mathbf{Q}^n_{i+1} - \mathbf{Q}^n_i}{\mathbf{Q}^n_i - \mathbf{Q}^n_{i-1}},\quad \mathbf{r}^R_{i+3/2} = \frac{\mathbf{Q}^n_{i+1} - \mathbf{Q}^n_i}{\mathbf{Q}^n_{i+2} - \mathbf{Q}^n_{i+1}} \tag{6a, b}$$

The first elements of the vectors $\mathbf{Q}^L$ and $\mathbf{Q}^R$ are updated as follows

$$\mathbf{Q}^L_{i+1/2}(1) = \phi[\mathbf{Q}^L_{i+1/2}(4)] + (1-\phi)[\mathbf{Q}^L_{i+1/2}(5) - z_{i+1/2}] \tag{7a}$$

$$\mathbf{Q}^R_{i+1/2}(1) = \phi[\mathbf{Q}^R_{i+1/2}(4)] + (1-\phi)[\mathbf{Q}^R_{i+1/2}(5) - z_{i+1/2}] \tag{7b}$$

where $z_{i+1/2} = (z_i + z_{i+1})/2$, the first term on the right hand side of Eq. (7a, b) represents the interpolated flow depth by the DGM, the second term on the right hand side of Eq. (7a, b) represents the one by the SGM, and $\phi$ = weighting factor between the DGM and the SGM with $0 \leq \phi \leq 1$. Here the parameter $\phi$ is specified as a function of the Froude number following Aureli et al. (2008)

$$\phi = \begin{cases} 0.5[1 - \cos(\frac{\pi Fr_{\lim}}{Fr})] & Fr > Fr_{\lim} \\ 1 & Fr < Fr_{\lim} \end{cases} \quad (8)$$

where $Fr_{\lim}$ is a critical Froude number below which the DGM version will be implemented. In this paper $Fr_{\lim} = 2.0$ is adopted.

The interpolated vectors $\mathbf{Q}^L$ and $\mathbf{Q}^R$ are used to update the variable vectors of the governing equations as follows

$$\mathbf{U}^L_{i+1/2} = [\mathbf{Q}^L_{i+1/2}(1), \mathbf{Q}^L_{i+1/2}(2), \mathbf{Q}^L_{i+1/2}(3)]^T \quad (9a)$$

$$\mathbf{U}^R_{i+1/2} = [\mathbf{Q}^R_{i+1/2}(1), \mathbf{Q}^R_{i+1/2}(2), \mathbf{Q}^R_{i+1/2}(3)]^T \quad (9b)$$

***Step 2:*** *Temporal evolution for the second-order accuracy*

The interpolated cell edge variables from the first step are further evolved over a half time step as

$$\overline{\mathbf{U}}^L_{i+1/2} = \mathbf{U}^L_{i+1/2} - \frac{\Delta t/2}{\Delta x}[\mathbf{F}(\mathbf{U}^L_{i+1/2}) - \mathbf{F}(\mathbf{U}^R_{i-1/2})] + \frac{\Delta t}{2}\mathbf{S}_{bi} \quad (10a)$$

$$\overline{\mathbf{U}}^R_{i+1/2} = \mathbf{U}^R_{i+1/2} - \frac{\Delta t/2}{\Delta x}[\mathbf{F}(\mathbf{U}^L_{i+3/2}) - \mathbf{F}(\mathbf{U}^R_{i+1/2})] + \frac{\Delta t}{2}\mathbf{S}_{bi+1} \quad (10b)$$

The second element of the vector of the bed slope source term in Eq. (10) is estimated as

$$\mathbf{S}_{bi}(2) = -g[\mathbf{U}^L_{i+1/2}(1) + \mathbf{U}^R_{i-1/2}(1)](z_{i+1/2} - z_{i-1/2})/(2\Delta x) \quad (11)$$

Here only the expression for the second element of the vector of the bed slope source term is given because the first and third elements are zero, see Eq. (2c).

***Step 3:*** *The numerical flux and the bed slope source term*

The numerical flux is estimated by the FORCE (first order centered) approximate Riemann solver, which is an average of the Lax-Friedrichs flux $\mathbf{F}^{LF}$ and the two-step Lax-Wendroff flux $\mathbf{F}^{LW2}$ (Toro *et al*. 2009)

$$\mathbf{F}_{i+1/2} = (\mathbf{F}^{LW}_{i+1/2} + \mathbf{F}^{LF}_{i+1/2})/2 \quad (12)$$

with

$$\mathbf{F}_{i+1/2}^{LW} = \mathbf{F}(\mathbf{U}_{i+1/2}^{LW}) \tag{13a}$$

$$\mathbf{U}_{i+1/2}^{LW} = \frac{1}{2}(\overline{\mathbf{U}}_{i+1/2}^{L} + \overline{\mathbf{U}}_{i+1/2}^{R}) - \frac{1}{2}\frac{\Delta t}{\Delta x}(\mathbf{F}(\overline{\mathbf{U}}_{i+1/2}^{R}) - \mathbf{F}(\overline{\mathbf{U}}_{i+1/2}^{L})) \tag{13b}$$

$$\mathbf{F}_{i+1/2}^{LF} = \frac{1}{2}(F(\overline{\mathbf{U}}_{i+1/2}^{L}) + F(\overline{\mathbf{U}}_{i+1/2}^{R})) - \frac{1}{2}\frac{\Delta x}{\Delta t}(\overline{\mathbf{U}}_{i+1/2}^{R} - \overline{\mathbf{U}}_{i+1/2}^{L}) \tag{13c}$$

Finally, the bed slope source term in Eq. (3) can be estimated in a similar way to the one in Eq. (10), but using the temporal evolved cell-edge variables, which reads

$$\mathbf{S}_{bi}(2) = -g[\overline{\mathbf{U}}_{i+1/2}^{L}(1) + \overline{\mathbf{U}}_{i-1/2}^{R}(1)](z_{i+1/2} - z_{i-1/2})/(2\Delta x) \tag{14}$$

*2.2.3 Wet/dry front*

The treatment of the wet/dry front is critical for the successful modeling of the swash flow since the swash water periodically runs up and down. A threshold flow depth $h_{\lim}$ is used to judge whether the cell is dry or wet. Two neighboring cells will be defined as the wet/dry front if one is wet and the other is dry. If the water level of the wet cell is higher than the bed elevation of the dry cell, the dry cell is going to become a wet cell; otherwise the inter-cell numerical flux is set to zero. The threshold flow depth, $h_{\lim}$, is a model parameter and should be sufficiently small for quantitative accuracy.

**3. Numerical case studies**

The performance of the present model is exhaustively demonstrated through five phases: 1) test its capability to model shallow water phenomena; 2) test against analytical solution without sediment transport; 3) test against quasi-analytical solution with sediment transport; 4) test against existing numerical solutions; and 5) test against experimental swash over a mobile bed.

**3.1 Capability to model shallow water phenomena**

The capability of the present model to track bore propagation (i.e., shock-capturing) and preserve the quiescent flow (i.e., the well-balanced property) are examined. Since there has been no standard case for testing the well-balanced property in the context of coastal flows (Brocchini and Dodd 2008), a unified numerical case is proposed here to examine both the well-balanced property and the ability to tracking bore propagation.

To test the well-balanced property of a model for the coastal flow, the topography must have two elements (Brocchini and Dodd 2008): 1) a shoreline, and 2) a change in the submerged beach slope. As shown in Fig. 1, a configuration fulfilling these requirements is designed. It is the same as one of the two popular experimental setups for investigating the swash flow (Kikkert et al. 2010; O'Donoghue et al. 2010), yet with an additional triangular hump. The configuration consists of three parts: on the left side is a reservoir, on the right ride is a sloping beach (bed slope: $\tan(\alpha)$, where $\alpha$ is the angle of the beach slope), and between them is a horizontal region with a symmetrical triangular hump. The following parameters should be prescribed: lengths of the reservoir part and the horizontal part, the base length and the height of the hump, the length between the reservoir and the hump, the length between the hump and the beach, the water height in and downstream the reservoir, the roughness, and the angle of the beach.

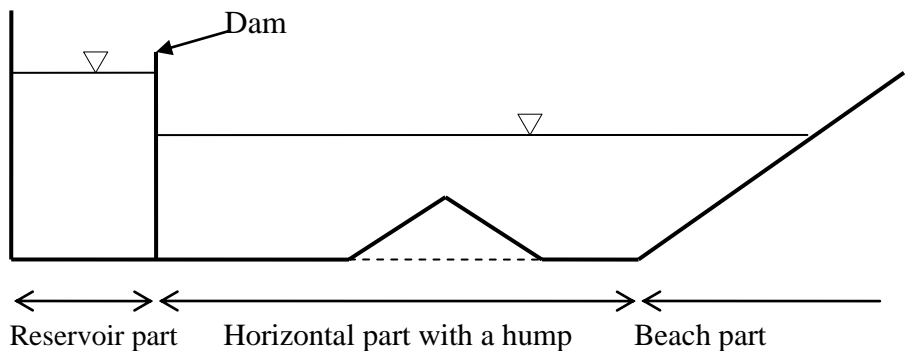

**Figure 1. Configuration for testing capabilities of model to track shallow water phenomena**

The first case with respect to Fig. 1 is used to test the ability to tracking bore propagation and described in the following. It approximates an experiment conducted by Ozmen-Cagatay et al. (2014): the length of the reservoir is 4.65 m; the triangular hump is symmetrical with a base length of 1 m and a height of 0.075 m; the center of the hump locates 2.0 m to the right of the dam, the angle $\alpha = 0$, the water depth in the reservoir is 0.25 m, and the bed is initially dry downstream. Manning roughness $n = 0.01$ because the flume bottom is built of glass. Sediment transport rate is set to zero. The spatial step $\Delta x = 0.01$ m and the Courant number $Cr = 0.1$, which are sufficiently small to assure grid-size-independence of the numerical results. This is also fulfilled for all test cases presented in this paper.

Fig. 2 presents the computed and measured longitudinal profiles of the water surface level in the experiment. The instant removal of the dam leads to collapse of the ponded water and forms a bore which propagates downstream. When the bore encounters the irregular hump, the water stage upstream of the hump increases due to the effect of reflection and a hydraulic jump occurs downstream of the hump (Fig. 2a). On the left side of the hydraulic jump, the increase of water stage expands further left and forms a second bore traveling backwards (Figs. 2a-2f). This leftwards-traveling bore would finally encounter the solid wall, which however is not shown due to lack of measurements. The computed water surface level agrees with the measurement very well around the hump. To further quantify the relative difference between the numerical solution and the measured data, the non-dimensional discrepancy for the flow depth is defined following the $L^1$ norm as $L^1_h = \sum |h_i - \hat{h}_i| / \sum h_i$ (Li et al. 2013) ($\hat{h}_i$ represents the measured and $h_i$ the corresponding computed flow depth). The non-dimensional discrepancies for the flow depth in relation to the comparisons shown in Fig. 2(a, b, c, d, e, f) are (0.0542, 0.0809, 0.081, 0.067, 0.082, 0.049), suggesting a good numerical accuracy of the model.

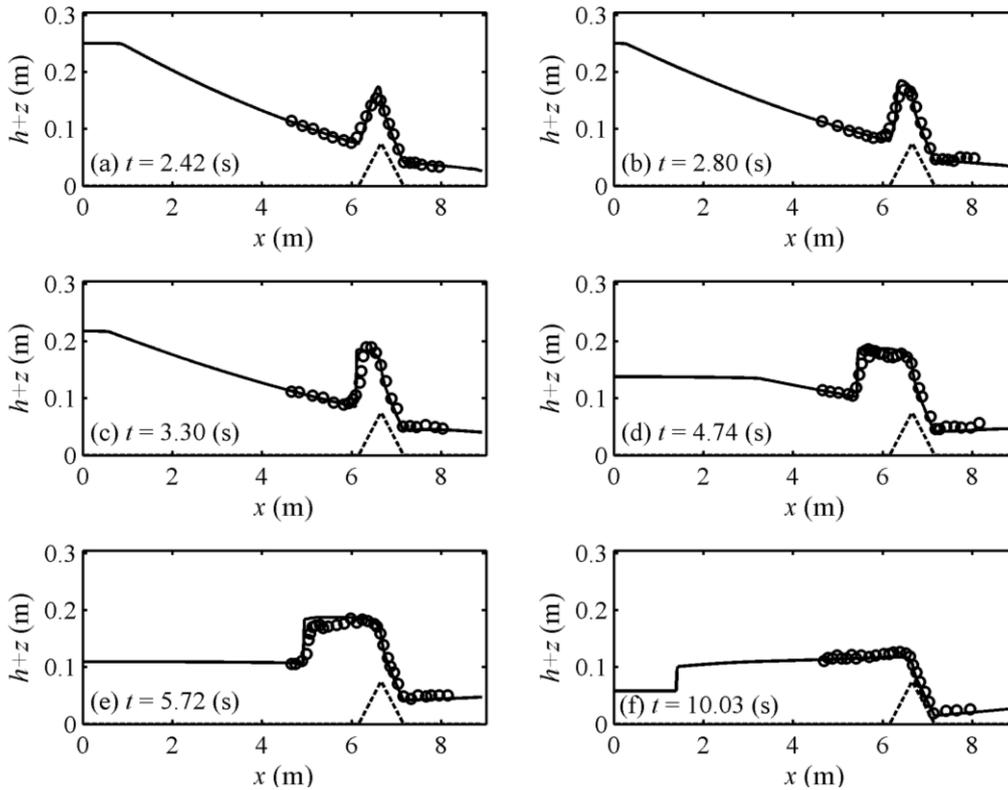

**Figure 2. Computed (solid line) and measured (dots) water surface level for the experiment of Ozmen-Cagatay et al. (2014).**

The second case is used to test the well-balanced property. It differs from the first case in two aspects: 1) the water depth in and downstream of the reservoir is set the same: 0.25 m; 2) the angle of the beach slope is not vanishing but set as $\alpha = 0.1$. These two aspects assure a quiescent flow.

Fig. 3(a, c) presents numerical results for this case computed by the well-balanced model. For comparison purposes, the numerical results computed from an un-balanced model (the same as the well-balanced model but with weighting factor $\phi = 1$) are shown in Fig. 3(b, d). While both the well-balanced model and the unbalanced model give a stable water depth, the computed flow velocity by the unbalanced model is appreciable (in the magnitude of $10^{-2}$ m/s), whereas those by the well-balanced model is negligible (in the magnitude of $10^{-15}$ m/s). This has important implications for coastal engineering modeling practices. Ignoring the well-balanced property

induces appreciable errors in modeling the quiescent flow in the coastal region especially for the flow velocity. Although coastal flows are mostly highly dynamic, it is unknown how the ignorance of the well-balanced property affects the computation of the coastal flows. Even though the uncertainty of this ignorance in modeling of some classical test cases appears limited (Dodd 1998; Hu *et al.* 2000; Wei *et al.* 2006; Brocchini *et al.* 2001; Briganti and Dodd 2009a, b; Mahdavi and Talebbeydokhti 2009, 2011; Kelly and Dodd 2010; O'Donoghue *et al.* 2010; Kuiry *et al.* 2012; Postacchini *et al.* 2012, 2014; Zhu *et al.* 2012; Zhu and Dodd 2013), the complexity of real processes may necessitate the consideration of the well-balanced property.

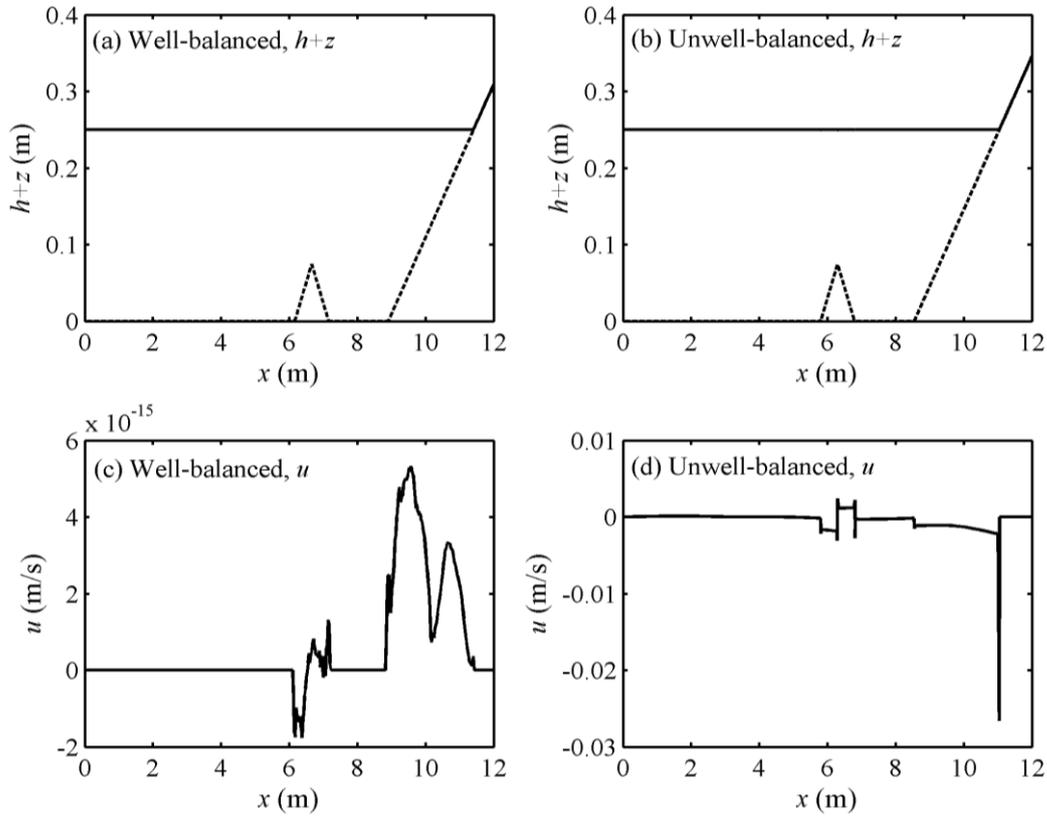

**Figure 3. Numerical results for simulating the quiescent flow from the well-balanced model and the unbalanced model.**

**3.2 Analytical solutions without sediment transport**

Here the analytical solutions of the SM63 (Shen and Meyer 1963) and the PW01 (Peregrine and Williams 2001) are numerically revisited by the present well-balanced model.

*3.2.1 SM63 and PW01 solutions*

The SM63 solutions, which represent the evolution of an incoming uniform bore on a beach of constant slope, read

$$h(x,t) = \frac{(2t\sqrt{g2H} - gt^2 \sin\alpha - 2x/\cos\alpha)^2}{36gt^2 \cos^2\alpha} \tag{15a}$$

$$u(x,t) = \frac{t\sqrt{g2H} - 2gt^2 \sin\alpha + 2x/\cos\alpha}{3t/\cos\alpha} \tag{15b}$$

where the parameter $\alpha$ represents the angle of the beach slope, $H = 2h_0 \cos\alpha$ corresponding to the maximum run-up height, and $h_0$ = water depth of the incoming bore. The toe of the swash zone is defined to be located at position $x = 0$. Theoretically, the location of vanishing water depth (i.e., $h = 0$ m) represents the shoreline position. Yet this definition may be practically difficult to operate with. Therefore, a notional shoreline position is defined using a critical flow depth $h_c$ that can be regarded as the minimum measurable water depth by a run-up probe. A similar practice can be found in Hubbard and Dodd (2002). By setting $h = h_c$, one can derive the expression for the notional shoreline position

$$x_s(t, h_c) = \frac{(4\sqrt{gA} - 6\sqrt{g\cos\alpha h_c})\cos\alpha t - gt^2 \sin\alpha \cos\alpha}{2} \tag{16}$$

Based on the SM63 solutions, Peregrine and Williams (2001) considered swash overtopping a truncated beach and obtained an analytical expression describing the volume $V$ of the overtopping flow against the non-dimensional cutoff position $E$ of the beach, which reads

$$V(E) = \frac{H^2}{54\sin(2\alpha)}(4 - 12E + 8E\sqrt{2E} - 3E^2) \tag{17}$$

This is referred to as the PW01 solution.

*3.2.2 Numerical results and analysis*

The SM63 and the PW01 solutions can be numerically solved as either an Initial Value Problem (IVP) or a Boundary Value Problem (BVP) (Pritchard and Hogg 2005). For both the IVP and BVP, a beach of constant slope [slope $= \tan\alpha$] is considered. For the BVP, only the portion of the swash zone (positions with $x \geq 0$) is considered in the computation. The position $x = 0$ is seen as the seaward boundary, where a boundary condition is provided (Guard and Baldock 2007): $u/\sqrt{gH/2} + 2\sqrt{2\cos\alpha h/H} + \sin\alpha \sqrt{2g/H} t = 2$. An initial water depth ($h_0$) must be fed into the boundary condition to obtain an initial flow velocity, so that the bore intensity is not arbitrary. For the IVP, both the swash zone and the seaside (i.e., positions with $x < 0$) are considered: water is initially at rest with a uniform water depth ($h_0$) at the seaside, and the swash zone is set as dry.

Numerical results of the IVP and the BVP are compared for a wide range of parameters, showing negligible difference between the two treatments. Thus, no reference is made to the IVP and BVP when presenting numerical results below. Also only the numerical results in the swash zone are presented, which correspond to the following parameters: the angle of the beach slope $\alpha = 0.1$, the initial water depth $h_0 = 0.6$ m, $q_s = 0$, and $n = 0$. Setting a zero value for the Manning's roughness coefficient means no friction from the bed. This is due to the fact that in the considered cases of the SM63 and the PW01, only solutions without friction are considered. This may be due to the complexity of the swash sediment transport that suggests a leading-order description in which friction is neglected (Brocchini and Baldock 2008). The spatial step $\Delta x = 0.005$ m and the Courant number $Cr = 0.1$.

Fig. 4 presents comparisons between the numerical prediction and the SM63 solution: a) water depth at different instants, b) flow velocity profiles, and c) the notional shoreline position. As seen in Fig. 4, the difference between the numerical results and the SM63 solution are negligible, meaning that the model satisfactorily reproduces the SM63 analytical solution. The detailed SM63 swash process is not repeated here for simplicity, as it has been extensively described previously (Shen and Meyer 1963; Peregrine and Williams 2001).

Fig. 5 shows a satisfactory agreement between the overtopping flow volume computed by the model and that estimated by the PW01 solution. The more the beach is truncated (smaller value for $E$), the larger is the volume of the overtopping flow. Each data point of the computed overtopping flow volume shown in Fig. 5 is estimated from the numerical results by integrating the product of water depth and velocity at the cut-off position over time.

The validation of the present model against the classical SM63 and PW01 solutions (Figs. 4 and 5) preliminarily indicates that the quantitative accuracy of the present model is high.

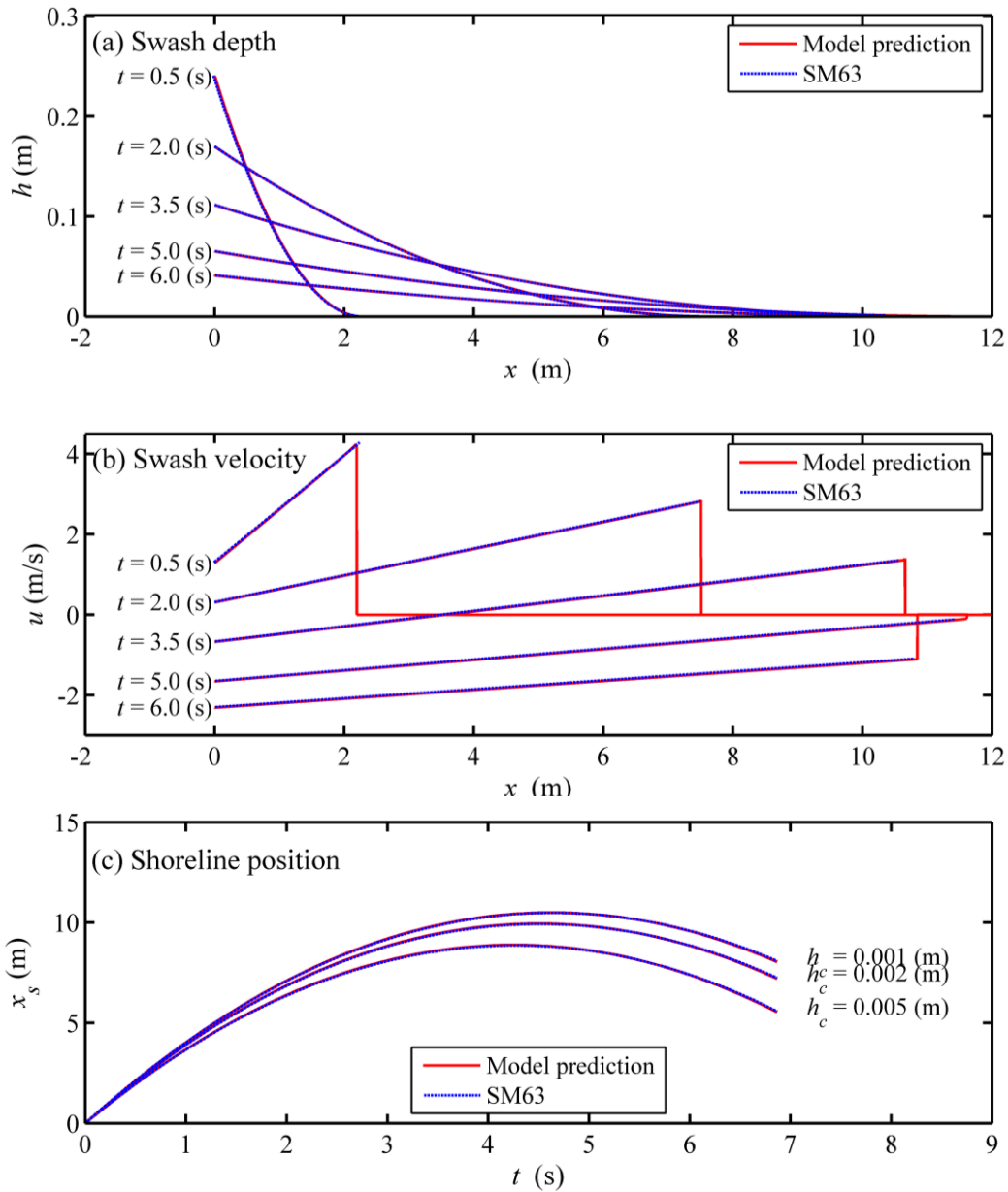

**Figure 4.** Comparisons between the numerical prediction and the SM63 solution: a) water depth, b) flow velocity, c) notional shoreline positions

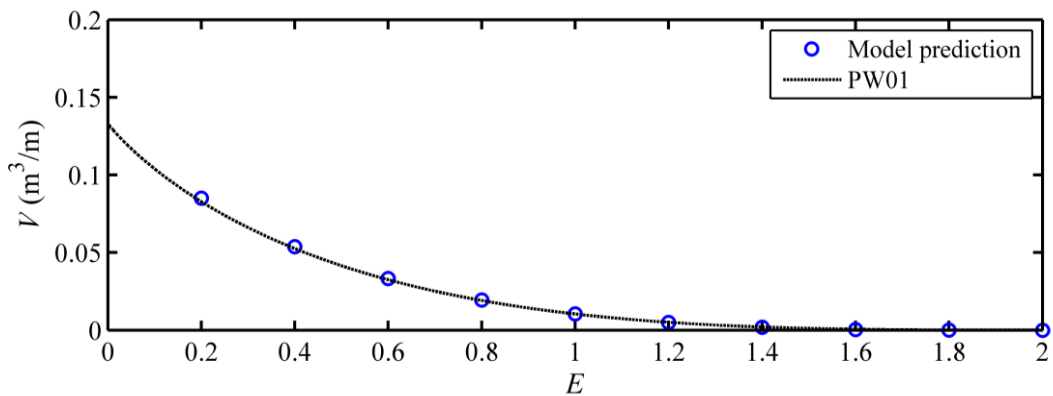

**Figure 5.** Overtopping water volume against truncated position: comparison between the computed value and the PW01 solution.

## 3.3 Quasi-analytical solutions with sediment transport

When sediment transport is involved, it is often challenging to obtain analytical solutions. Nonetheless, an approach of obtaining quasi-analytical solutions for dam-break flow with sediment transport has been proposed by Kelly (2009) and Kelly and Dodd (2009), which is referred to KD09 here. The KD09 solutions can be obtained by iterating procedures and have been widely used for testing of coastal numerical models (Kelly and Dodd 2010, Briganti et al. 2012a, Postacchini et al. 2012) due to the fact that dam-break flows represent a good IVP interpretation of the BVP of coastal flows (Kelly 2009). For details on the procedures of obtaining the KD09 solutions, one can refer to Kelly (2009) and Kelly and Dodd (2009, 2010).

Here KD09 solutions are compared with numerical results computed by the present model. Two cases are considered with different bed mobility, as manifested in the empirical sediment transport relation: one uses $q_b = 0.004 u^3$ representing a mobile bed, and the other uses $q_b = 10^{-8} u^3$ representing essentially a fixed bed. The initial conditions are as follows: for $x \leq 0$, the water is at rest and has a uniform depth of $h_0 = 1$ m, and for $x > 0$, the bed is dry with a vanishing water depth. The flow is initiated by instant removal of an assumed dam at $x = 0$. The spatial step $\Delta x = 0.005$ m and the Courant number $Cr = 0.1$.

Figs. 6 and 7 present the comparisons between the numerical results and the KD09 solutions: longitudinal profiles of water surface level, bed elevation and non-dimensional flow velocity at five instants. Fig. 6 correspond to $q_s = 0.004 u^3$, whereas Fig. 7 correspond to $q_s = 10^{-8} u^3$. From Figs. 6 and 7, the KD09 solutions are satisfactorily reproduced by the model: the numerical results and the KD09 solutions almost overlap perfectly over the whole computational domain, which is similar to the model of Postacchini et al. (2012) but much improved compared to those of Briganti et al. (2012a), suggesting that the present model provides a good description of this challenging dam-break evolution over a mobile bed.

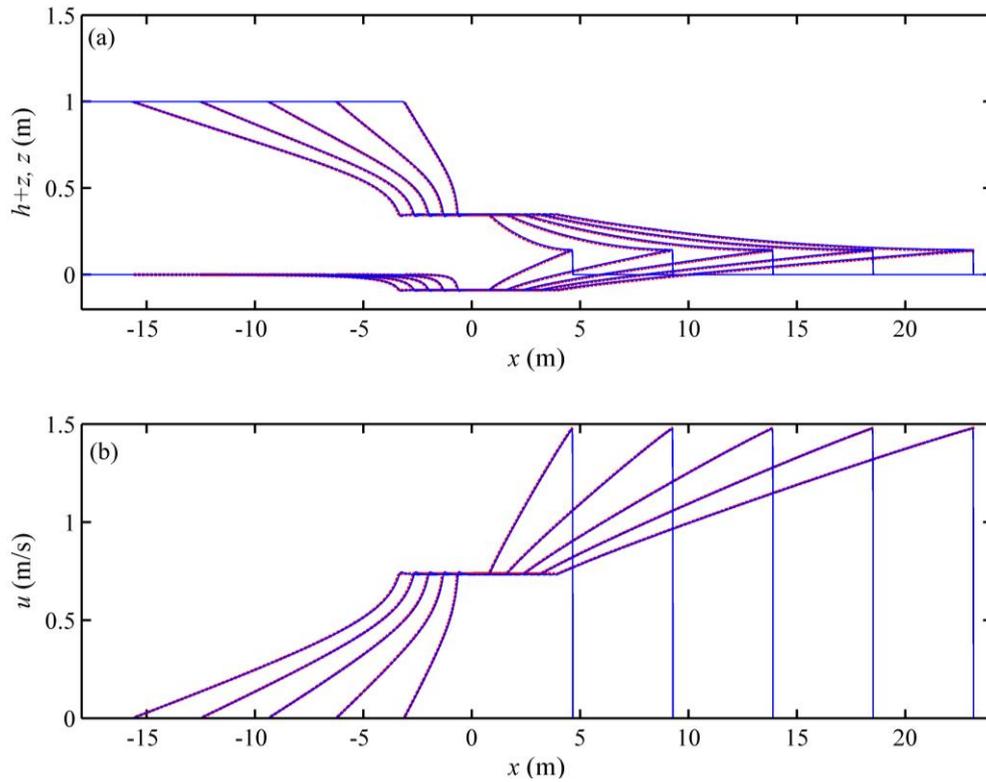

**Figure 6.** Comparisons between the numerical results (solid blue lines) and the KD09 (dashed red lines) solutions for $q_s = 0.004u^3$: longitudinal profiles of a) water surface level and bed elevation, and b) flow velocity at five instants ($t$ = 1s, 2s, 3s, 4s and 5s).

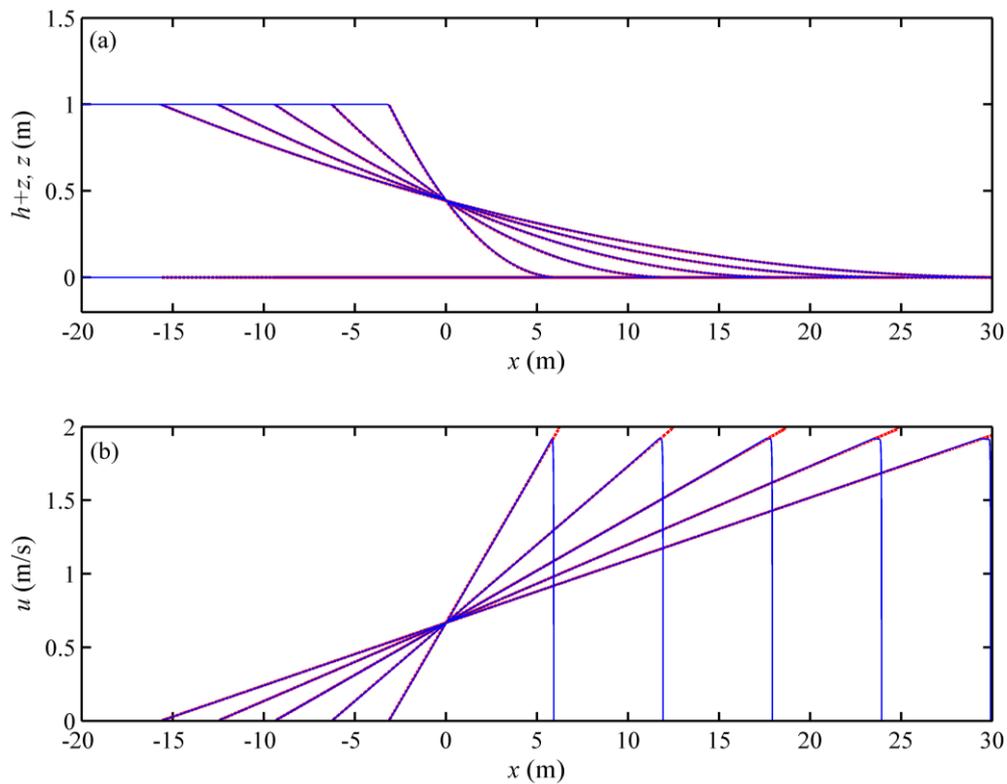

**Figure 7.** Comparisons between the numerical results (solid blue lines) and the KD09 (dashed red lines) solutions for $q_s = 10^{-8}u^3$: longitudinal profiles of a) water surface level, and bed elevation, and b) flow velocity at five instants ($t$ = 1s, 2s, 3s, 4s and 5s).

## 3.4 Existing numerical solutions for swash flow over a mobile bed

Here existing numerical solutions are deployed to test the present model. To this end, the existing numerical solutions must be of high accuracy. Fortunately, those by Zhu and Dodd (2013) fulfill this requirement because they are obtained with the specified time interval method of characteristics (STI MOC). In Zhu and Dodd (2013), beach morphological evolutions in response to the SM63/PW01 swash event are numerically investigated as the IVP, for which a beach with an initial bed slope of 0.1 is considered, with the position $x = 0$ m denoting the toe of the swash zone. The initial conditions are as follows: for positions with $x \leq 0$, the water is at rest and has a uniform depth of $h_0 = 0.65$ m, and for the swash zone with $x > 0$, the bed is dry with vanishing water. Five empirical sediment transport relations are used, including the Grass formula: $q_b = A_G u^3$, the Bagnold formula: $q_b = A_{Bn} u(u^2 - u_e^2)$, the MPM formula: [$q_b = A_M (u^2 - u_e^2)^{3/2}$, the van Rijn formula: $q_b = A_v u |u|^{2.4}$, and the Bailard formula : $q_b = A_{Bl} u |u|^3$, where $A_G$, $A_{Bn}$, $A_M$, $A_v$, $A_{Bl}$ are coefficients, and $u_e$ is the threshold velocity for the incipient sediment motion. Note that the implementation of the five empirical sediment transport relations in the model by Zhu and Dodd (2013) does not imply the Zhu and Dodd (2013) model is flexible. Rather, the implementation of these empirical sediment transport relations is achieved as the cost of much additional efforts. These coefficients have been chosen to assure the same net offshore sediment flux (about 0.4457) at the toe of the swash zone ($x = 0$), which result in $A_G$ =6.1224E-4, $A_{Bn}$ =7.2E-4, $A_M$ =7.7143E-4, $A_v$ =3.7541E-4, $A_{Bl}$ =1.6811E-4, and $u_e$ = 1.1358 m/s. These numerical cases are revisited by the present model as the IVP. Numerical solutions by Zhu and Dodd (2013) are referred to as ZD13. The spatial step $\Delta x = 0.005$ m and the Courant number $Cr = 0.1$.

Fig. 8 presents the distribution in the space-time plane of the beach deformation depth when the Grass formula is used to compute the sediment transport rate by the present model (Fig. 8a) and by the ZD13 (Fig. 8b). Table 1 presents a summary of the beach deformation depth at the toe of the swash zone in relation to the five empirical sediment relations. It is obvious from Fig. 8 that the present model and the ZD13 give almost the same beach morphological deformation. Furthermore, the quantitative comparisons presented in Table 1 suggest a high accuracy of the present model. In fact, the computed non-dimensional discrepancy $L^1_{\Delta z} = \sum |\Delta z_i - \Delta \hat{z}_i| / \sum \Delta z_i$, where $\Delta z_i$ represents the beach deformation depth computed by the present model and $\Delta \hat{z}_i$ the beach deformation depth computed by ZD13 are within about 1% for all the five empirical sediment transport relations).

It is noted that for these cases, only the computed beach morphological change is compared. This is reasonable because the effect of a change in the beach profile has been immediately fed to the swash hydrodynamics. It is worthy noting that the MOC (i.e., the ZD13) has a relatively low flexibility in dealing with sediment transport and a high complexity of the model structure due to the involvement of the eigenstructure. These indicate that the present model has maintained the numerical accuracy, improved the flexibility, and at the same time reduced model complexity, which is useful for coastal engineering purposes.

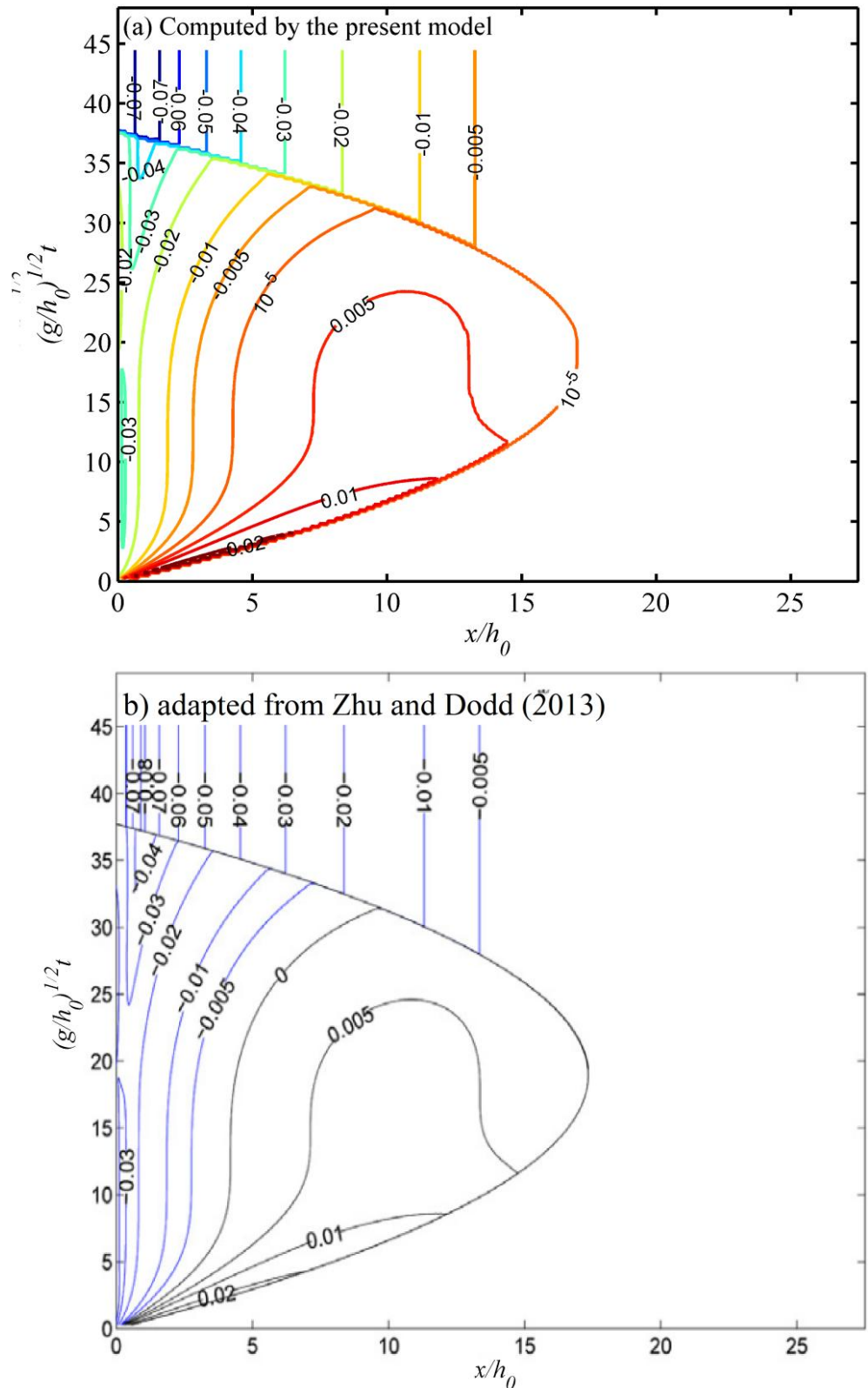

**Figure 8.** Distributions in the space-time plane of the computed beach deformation depth when the Grass formula is used to compute the sediment transport rate by a) the present model, and b) ZD13.

**Table 1. Computed non-dimensional beach deformation depth at the toe of the swash zone**

|  | Grass | Bagnold | MPM | Van Rijn | Bailard |
|---|---|---|---|---|---|

| ZD13 | -0.0601 | -0.0604 | -0.0615 | -0.0639 | -0.0707 |
| The present model | -0.0603 | -0.0604 | -0.0618 | -0.0646 | -0.0701 |
| $L^1_{\Delta z}\ (=\sum|\Delta z_i - \Delta \hat{z}_i|/\sum \Delta z_i)$ | **0.33%** | **0.00%** | **0.49%** | **1.08%** | **0.85%** |

### 3.5 Experimental swash without/with sediment transport

Two categories of experimental configurations have been used to study the swash flow previously: one makes use of the dam break flow to simulate the bore/wave (O'Donoghue *et al.* 2010; Kikkert *et al.* 2010); the other one uses the wave-maker to initiate the bore/wave (Baldock *et al.* 2005). However, most experiments focus on the swash flow free of sediment transport. Here the experimental dam-break driven swash over a mobile bed by Kikkert *et al.* (2010) is numerically revisited.

The experimental setup is the same as that shown in Fig. 1, but without the triangular hump. The length of the reservoir part: 1.0 m; the beach slope: $\tan\alpha = 0.1$; the length of the horizontal part: 4.0 m; the water depth in the reservoir: 0.65 m; and the water depth downstream of the reservoir: 0.06 m. The position $x = 0$ m is defined at the initial shoreline position, and the x-axis is directed shoreward. The bed is composed of sediment particles with diameter $d_{50}= 1.3$ mm. The Manning roughness is estimated by $n = d_{50}^{1/6}/20$, where $d_{50}$ is in meter. The Grass formula $q_b = A_G u^3$ is used for the sediment transport rate, where $A_G = 0.0004\,\mathrm{m^2/s}$ following the calibration by Briganti *et al.* (2012b). All three parts (the reservoir, the horizontal region, and the beach) are considered in the computation. The spatial step $\Delta x = 0.005$ m and the Courant number $Cr = 0.1$.

Fig. 9 presents the computed and measured time variations of the swash water depth (Fig. 9a, c, e) and the swash velocity (Fig. 9b, d, f) at the three positions ($x = $ 0.072 m, 0.77 m and 2.36 m) along the beach. It is seen from Fig. 9 that the computed swash depth and velocity agree with the measured data quite well. Only when small flow depth dominates [i.e., in the cases of short instants when bore recedes (Fig. 9a, b, c, d), or at the position far from the initial shoreline position ($x = $ 2.36 m, see Fig. 9e, f)], discrepancies between the computation and the measurements are appreciable. This may be attributed to the effects of the very small flow depth that makes the use of a constant Manning coefficient less applicable. Using a constant Manning roughness and thus estimating the bottom shear stress as a bulk frictional force is due to the fact that a satisfactory relation for the friction of the swash zone is still missing (Hughes 1994; Puleo and Holland 2001), whereas using the bulk frictional force has been generally successful even for oscillatory motions (Brocchini et al. 2001). In this regard, it is noted that this experiment has also been numerically investigated by Briganti et al. (2012b) yet using a much more complex way of estimating the bottom shear stress, i.e., the momentum integral method. Nonetheless, using the complex momentum integral method does not appear to yield better results than the present practice: values of the $L^1_{q_b}$ norm in relation to the momentum integral method are (130%, 118%, 91%) for the three positions along the beach, whereas they are (28%, 30%, 55%) for the present model. Not surprisingly, the present model satisfactorily resolves the different magnitude of the sediment transport rate between the uprush phase and the backwash phase (Fig. 10), whereas this is missed by the momentum integral method. This is understandable from the governing equations: the sediment transport rate depends directly on the flow strength. While similar magnitudes of flow velocity between the uprush and backwash phases were computed by Briganti et al. (2012b), the computed backwash flow velocity is greatly reduced compared to that of the uprush, which is consistent with the measurements. Using the simple bulk frictional force to represent the bottom shear stress, the present model satisfactorily resolves this experimental swash flow over a mobile bed, suggesting a good accuracy of the present model.

## 4. Conclusions

This paper presents a well-balanced and flexible morphological modeling technique for the swash zone, which is motivated by two reasons: 1) the flexibility in dealing with sediment transport of existing morphological numerical models for the swash zone is limited, and 2) the well-balanced property of a model is rarely considered in previous morphological modeling of the swash flow. Both aspects are important for coastal engineering purposes. The high flexibility of the present model in dealing with sediment transport is achieved through using the finite volume method to discretize the governing equations and using the slope limited centered (SLIC) scheme to compute the numerical fluxes. The satisfaction of the well-balanced property is achieved by using the weighted surface depth gradient method (WSDGM) in the data reconstruction step of the SLIC scheme. The high flexibility is obvious from the fact the SLIC scheme avoids the use of the eigenstructure of the governing equations and is also demonstrated by the numerical reproduction of existing numerical results in relation to five empirical sediment transport relations. The satisfaction of the well-balanced property is demonstrated by numerical reproduction of a quiescent flow over an irregular bed that has two elements of coastal flows: a shoreline and a change in the submerged beach slope. The capabilities and quantitative accuracy of the model are shown to be improved or be at least at a similar level when compared to analytical solutions, existing numerical solutions of high accuracy, and experimental data. This work provides an improved modeling framework for swash hydrodynamics and sediment transport.

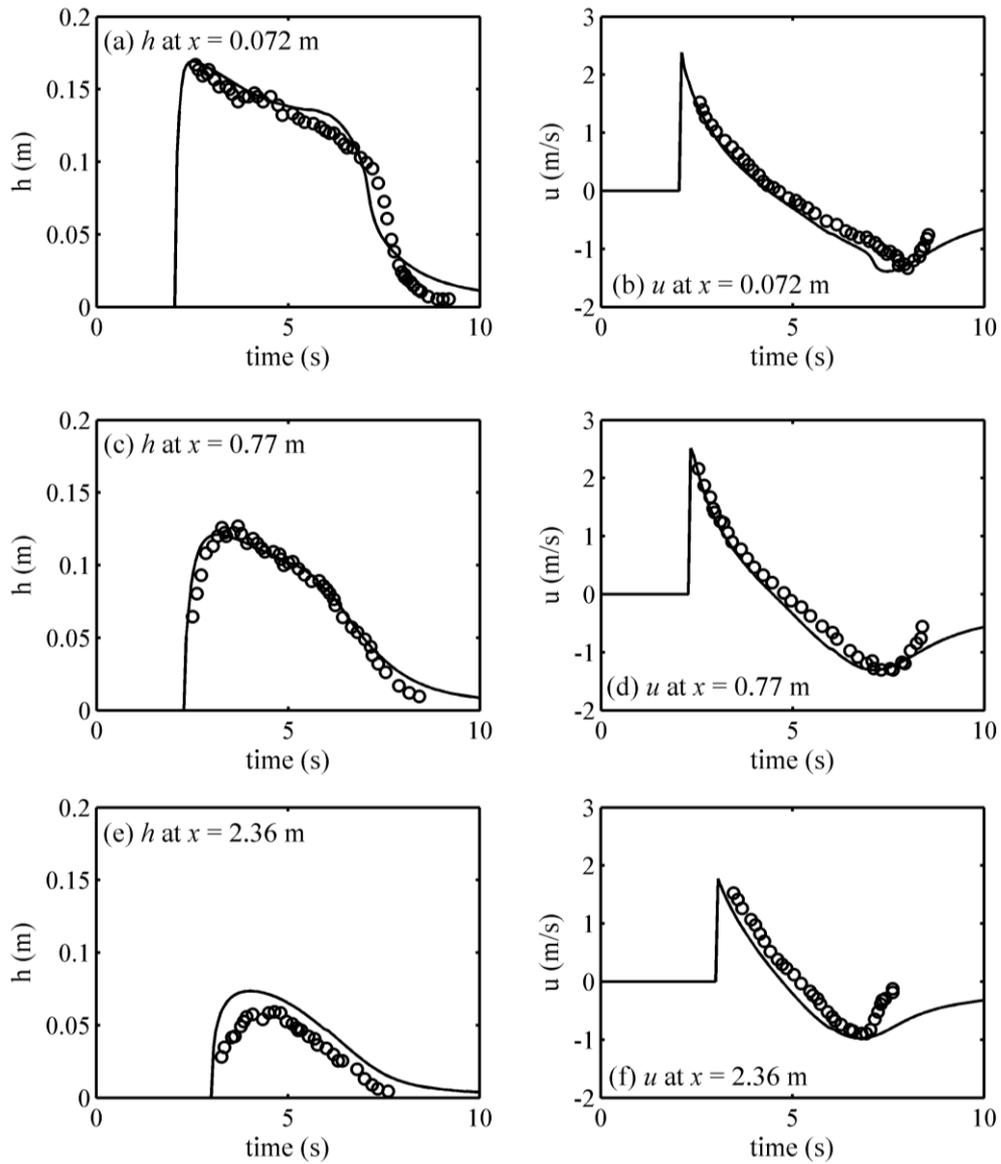

**Figure 9.** Comparisons between the numerical prediction and the experimental data: time variation of (a, c, e) the swash depth and (b, d, f) the flow velocity for the swash with sediment transport.

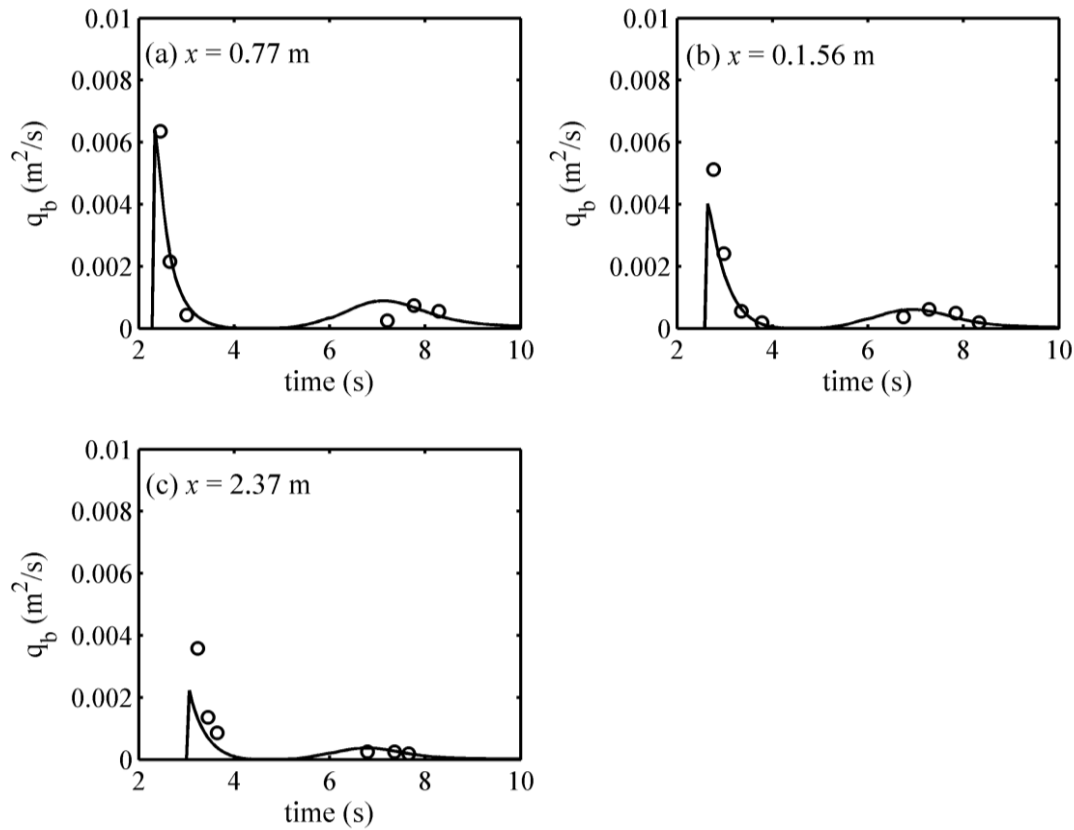

Figure 10. Comparisons between the numerical prediction and the experimental data: time variation of the sediment transport rate for the swash with sediment transport.

**Acknowledgements**

This research is supported by the Research Fund for Doctoral Program of Higher Education of China (Grant No. 20130101120152), the Fundamental Research Funds for Central Universities of China (Grant No. 2013QNA4041) and the National Natural Science Foundation of China (Grant Nos. 11402231, 41376095 and 41350110226).